\documentclass[printer]{aa} 
\usepackage{graphicx}
\usepackage{amsmath}
\usepackage{txfonts}
\usepackage{csquotes}
\MakeOuterQuote{"}

\begin{document} 

\title{On the Decades-Long Stability of the Interstellar Wind through the Solar System}

\author{R. Lallement
          \inst{1}
         \and
         J.L. Bertaux
         \inst{2}
          }

   \institute{GEPI Observatoire de Paris, CNRS, Universit\'e Paris Diderot, Place Jules Janssen  92190 Meudon, France\\
              \email{rosine.lallement@obspm.fr}
        \and
        LATMOS, Universit\'e de Versailles Saint Quentin, INSU/CNRS, 11 Bd D' Alembert, 78200 Guyancourt, France
             }

   \date{Received ; accepted }

 
  \abstract
 {We have revisited the series of observations recently used to infer a temporal variation of the interstellar helium flow over the last forty years. Concerning the recent  IBEX-Lo direct detection of Helium neutrals, there are two types of precise and unambiguous measurements which do not rely on the exact response of the instrument: the count rate maxima as a function of the spin angle, which determines the ecliptic latitude of the flow, and the count rate maxima as a function of IBEX longitude, which determines a tight relationship between the ecliptic longitude of the flow and its velocity far from the Sun. These measurements provide parameters (and couples of parameters in the second case) remarkably similar to the \textit{canonical, old} values. In contrast, the preferential choice of a lower velocity
and higher longitude reported before from IBEX data is based only on the count rate variation (at each spin phase maximum) as a function of the satellite longitude, when drifting across the region of high fluxes. We have examined the consequences of dead time counting effects, and conclude that their inclusion at a realistic level is sufficient to reconcile the data with the \textit{old} parameters, calling for further investigations. We discuss the analyses of the STEREO pickup ion data and argue that the statistical method that has been preferred to infer the neutral flow longitude (instead of the more direct method based on the pickup ion maximum flux directions), is not appropriate. Moreover, transport effects may have been significant at the very weak solar activity level of 2007-2009, in which case the longitudes of the pickup ion maxima are only upper limits on the flow longitude. Finally, we found that the use of some flow longitude determinations based on UV glow data are not adequate. Based on this global study, and at variance with recent conclusions we find no evidence for a temporal variability of the interstellar helium flow. This has implications for inner and outer heliosphere studies.}


   \maketitle
%

\section{Introduction}

The low energy, neutral helium flow measured in the solar system has its origin in the $\simeq$25 km.s$^{-1}$  relative motion of the Sun with respect to the ambient interstellar cloud. Interstellar helium measurements are crucial because, at variance with hydrogen, helium atoms experience very little charge-exchange with the solar wind ions, and as a consequence the helium flow velocity vector is the best indicator of the actual relative motion. The precise direction of the helium flow also serves as a reference for measurements of other species, and inferred differences can be used to derive the detailed coupling between the interstellar and solar plasmas (e.g. \cite{izmo99,lall05}. Possessing a very accurate value of the interstellar flow parameters is also particularly timely, because the two Voyager spacecraft are currently producing new and exciting data on the structure of the inner and outer heliosheath, including the recent crossing of the heliopause by Voyager 1 \citep{gurnett13,stone13,krimigis13}. 

After the pioneering measurements of the HeI 58.4nm solar  backscattered radiation \citep{weller74,ajello78,ajello79,weller79,weller81}, the first  \textit{in situ} measurements of interstellar helium were obtained, starting with pickup ions \citep{mobius85,mobius95} and then with direct neutral helium atom detection \citep{witte93}. Extensive efforts were done in 2003-2004 (ISSI workshop) to analyze in a synthesized manner all available datasets. In particular, the new analyses resolved the remaining strong discrepancies between flow temperature and velocities that were independently derived from UV and particle data \citep{lall04a,lalluvcs}. A canonical set of parameters, i.e. flow direction, flow speed, and temperature, was presented by \cite{mobius04} based on a synthesis of all data. Those flow parameters are very close to the parameters of \cite{witte04} derived from the Ulysses atom detection data. 

More recently, the Interstellar Boundary EXplorer (IBEX) mission has produced unprecedented measurements of neutral atoms in the heliosphere, in particular energetic atoms generated in the outer heliosphere \citep{mccomasribbon} as well as low energy atoms of direct interstellar origin, especially helium atoms \citep{hlond12}. The analysis of the first two years  of IBEX-Lo helium data (2009-2010) led to the derivation of helium flow parameters quite different  from the \textit{canonical} values, in particular a flow longitude of 79.2$^{\circ}$, instead of 75.4$^{\circ}$, and a speed of 22.8 km.s$^{-1}$ instead of 26.3$\pm$0.5 km.s$^{-1}$  \citep{bzozo12,mobius12}. Soon after those new parameters were published, \cite{drews12} announced similar values of the flow longitude, based on their analysis of pickup ions (PUIs) detected by STEREO/PLASTIC. 

The value of the flow velocity has a strong influence on the heliospheric interface structure, and indeed the IBEX-Lo low speed was used as an argument to dismiss the existence of an interstellar  bow-shock ahead of the heliopause \citep{mccomasbowshock}. It also motivated new theoretical studies about the actual nature of the outer heliosheath \citep{zank13,zieger13}. A compilation of all results obtained since the 70's and including the new IBEX-Lo and STEREO determinations has been interpreted as a statistically significant demonstration of a temporal change in the longitude of the interstellar helium flow velocity vector over the last decades \cite{crim13}, which is very surprising owing to the large mean free paths against collisions in low density clouds such as the circumsolar one. 

In this work we revisit the most recent datasets on the helium flow and their interpretations. Section 2 is devoted to the IBEX-Lo results that have a very strong statistical weight in the analysis of the longitude temporal variation, due to the accuracy of the measurements. In section 3 we discuss the other data, i.e. PUIs and backscattered radiation measurements. Section 4 synthesizes the data and most reliable parameters and discusses the likelihood of temporal variations of the flow over the last decades.

 \begin{figure*}
   \centering
\includegraphics[width=\linewidth]{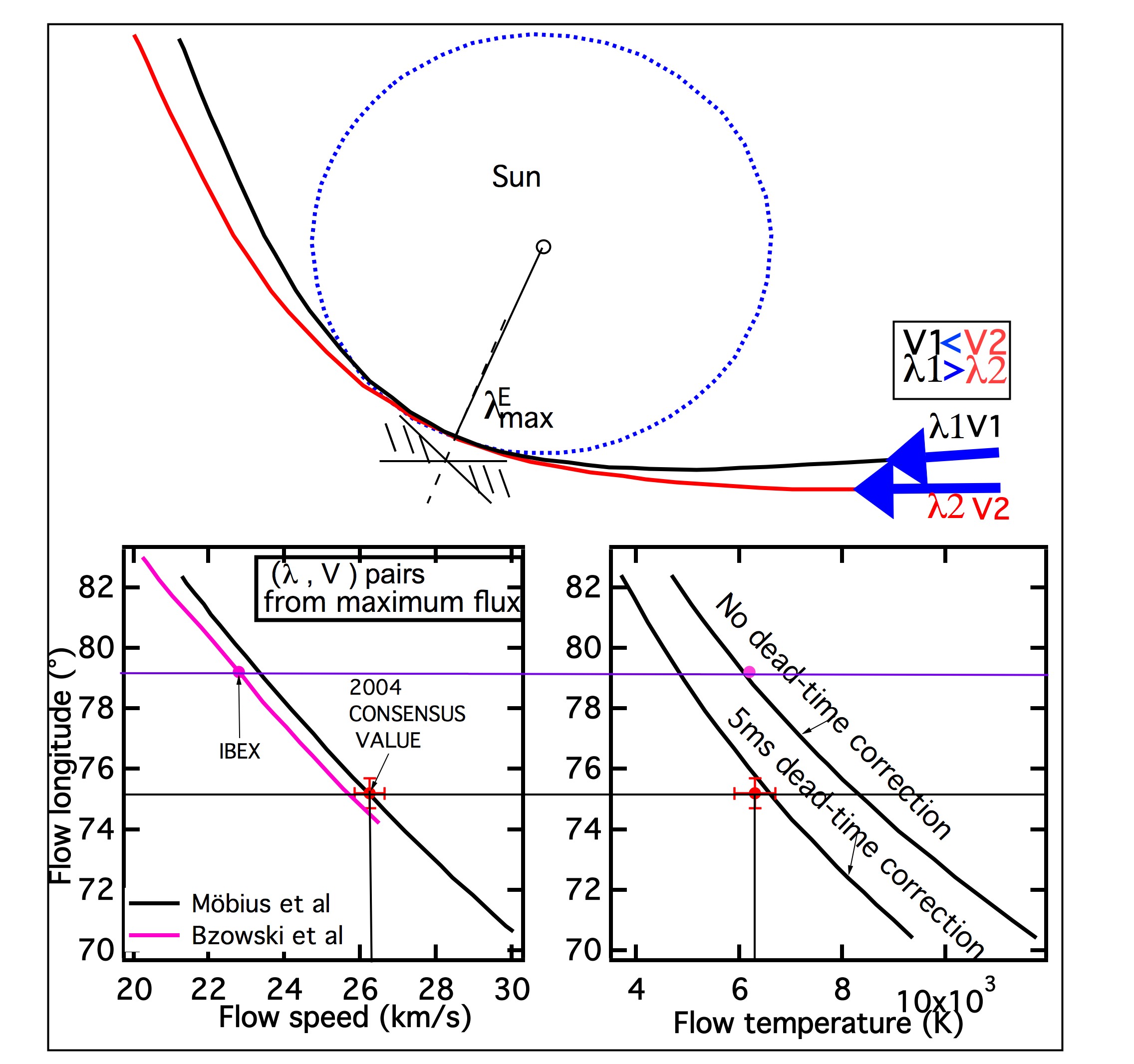}
   \caption{Top: An extremely solid measurement of IBEX-Lo is the Earth longitude of the maximum neutral atom flux $\lambda_{E}$, where the entrance field (hatched cones) is directed against the flow ($\simeq$ ortho-radial bulk flow). Each value of $\lambda_{E}$ corresponds to a continuum of pairs of flow longitude and speed values, as illustrated by the two (black and red) trajectories. Bottom left: the corresponding longitude-speed relationship inferred by \cite{mobius12} (black line) and based on the measurement of $\lambda_{E}$ with IBEX-Lo. Also shown is the central part of the chi-square parametric image of \cite{bzozo12}, that also corresponds to the same longitude-speed link. The \textit{consensus} pair of values is fully compatible with those relationships. Bottom right: the temperature-longitude derived by \cite{mobius12} based on the flux measurements during the rotation around the IBEX spin axis. The authors considered both the absence (top curve) and the presence (bottom curve) of high-rate suppression of events (or dead-time). The \textit{consensus} temperature of $\simeq$ 6,300K is compatible with the \textit{consensus} flow longitude of $\simeq$ 75.4 $^{\circ}$ for a dead time on the order of 5ms or slightly higher, i.e. realistic values according to \cite{mobius12}.}
              \label{figcurve}%
    \end{figure*}

  \section{Helium flow parameters from IBEX-Lo \textit{in situ} data}
With the IBEX satellite for the first time long duration direct measurements of helium atoms have been and are currently made from the Earth orbit \citep{mccomas12,hlond12}, in the keV energy range with IBEX-Hi \citep{IBEXHI09} and at lower energy with IBEX-Lo \citep{IBEXLO09}. Because gravitational effects and ionization by the solar radiation and solar particles have very strong impacts on the low energy interstellar flow within a few A.U., IBEX-Lo data are analyzed by comparison with models that follow the flow and take all those effects into account. \cite{bzozo12,mobius12} and \cite{lee12} have developed appropriate extensive analytical and numerical models supporting the analysis of the neutral helium data recorded since 2009. \cite{bzozo12} derived from the numerical analysis of the first two years of data, a set of parameters for the helium flow, in rough agreement with the more analytical determinations by \cite{mobius12}.

There are two puzzling coincidences in the \textit{new} IBEX-Lo helium parameters that have not been discussed and in our opinion deserve more attention. First, the pair of \textit{new} flow longitude $\lambda$ and velocity V values (at infinity) corresponds to a location of maximum count rate along the Earth orbit  that is remarkably similar to the location one would find for the \textit{old} parameters. This is important because an extremely solid and precise measurement of IBEX-Lo is indeed this Earth longitude of maximum count rate, that corresponds to the strongest flux of neutral atoms entering the IBEX-Lo field-of-view (FOV). This happens where the boresight axis is directed against the flow. Since this axis is close to perpendicular to the Sun-Earth direction this implies a helium flow that is close to tangential to the Earth motion,  or an ortho-radial bulk flow at 1 AU. It is well known however that  there is a continuum of flow longitude and speed pairs that correspond to the same unique value of Earth longitude $\lambda_{E}$ with such an orthoradial flow orientation, as illustrated by Fig \ref{figcurve}. The lower the flow speed at large distance from the Sun, the stronger the curvature of the trajectories due to the solar gravitational field, an effect that tends to increase $\lambda_{E}$ if the distant flow longitude remains the same, but can be compensated by an increase in the flow longitude that has the effect of rotating the whole set of trajectories around the Sun along an ecliptic polar axis. This longitude-speed relationship has been calculated analytically by \cite{mobius12}  and, independently, is found numerically by \cite{bzozo12} as the locus of minimum $\chi^{2}$ in the V-$\lambda$ space. Both (and very similar) relationships  taken from their respective publications and based on the precise measurement of $\lambda_{E}$ with IBEX-Lo, are displayed together in Fig \ref{figcurve}. It can be seen from the Figure how the pair of \textit{old}  values is quite compatible with this relationship. It implies that, if indeed the flow has rotated by about 4 degrees with respect to its former direction, coincidentally its speed has varied by the exact amount that is necessary to keep the same Earth location of tangential flow. Because the IS flow change is a phenomenon in the interstellar (IS) medium that is not related to the Earth orbit, this coincidence alone is  suspicious in itself. 

The second coincidence is the total absence of change of the flow ecliptic latitude $\beta$. As a matter of fact, while $\lambda$ has varied by about four degrees, remarkably no variation of  $\beta$ is detected, despite uncertainties on the latitude measurements that are far smaller (on the order of 0.2$^{\circ}$) than the uncertainties on the longitude. Interestingly another extremely solid measurement by IBEX-Lo is indeed this latitude $\beta$, accurately measured by the spin phase of the maximum count rate during the rotation of IBEX around an axis near the Sun direction. As a matter of fact, the latitude of this maximum count location changes very little under the effect of gravitation and Earth location. Here, again, the orientation of the ecliptic Plane and the interstellar medium spatial variations are unrelated and simultaneous changes of both the flow longitude and latitude are more likely than the change of only one of the two. We emphasize again that these two coincidences correspond to unambiguous measurements based only on {\bf maxima of the count rate}.  In both cases the measured quantities, namely the parameter $\beta$ and the combination of the two other parameters $\lambda$ and V correspond exactly to the previous \textit{consensus} values. 
Given these coincidences it is important to understand what is  influencing the derivation of the \textit{new} pair of parameters (79.2$^{\circ}$, 22.8 km.s$^{-1}$) that comes out from the works quoted above, and what dismisses the \textit{old} pair (75.4$^{\circ}$, 26.3 km.s$^{-1}$), knowing that they both correspond to the same Earth longitude for maximum rate. Since we fully trust the models carefully developed and cross-checked by \cite{lee12,mobius12,bzozo12}, a search should be oriented toward potential biases. Indeed, the existence of a bias is suspected, owing to the puzzling iso-$\chi^{2}$ curves presented by \cite{bzozo12}, that display two unexplained maxima. 

\begin{figure*}
   \centering
\includegraphics[width=\linewidth]{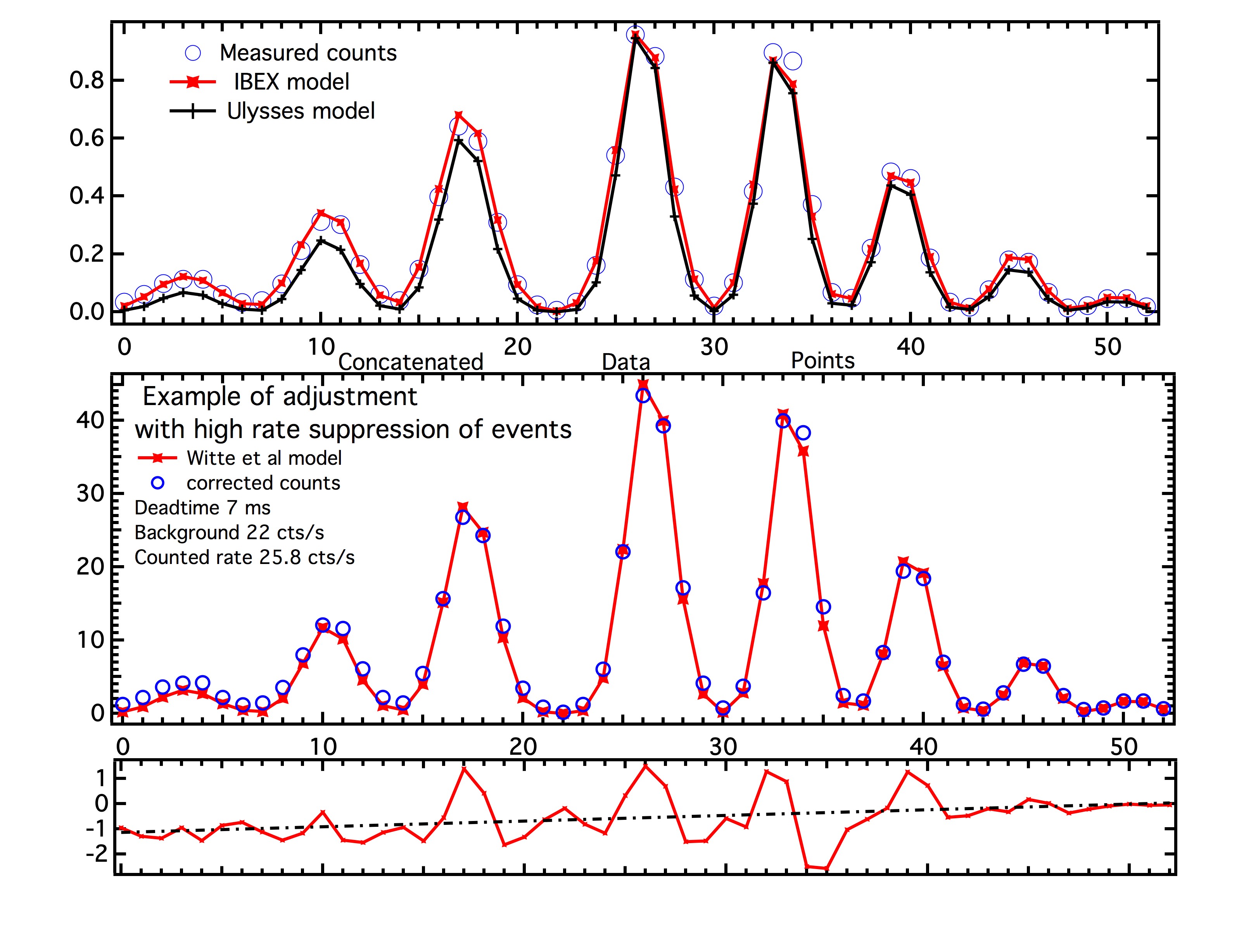}
   \caption{Top: Concatenated data and model values taken from \cite{bzozo12} for orbits 13 to 20 in year 2009. Both data and model have been precisely computed by the authors for the same geometry and the same time intervals. The abscissa represents the data point number and both data and models are the normalized values of \cite{bzozo12}. The Witte et al. model decreases more rapidly than the count rate for Earth ecliptic longitudes that are both lower and higher  than the maximum flux longitude $\lambda_{E}$. Middle: Example of scaling of the Witte et al. model to the count rates assuming a dead-time of 7 ms, a maximum count rate of 25.8 counts/s due to He atoms and an electron background of 22 counts/s. Units are counts/s. The sum of squared residuals after normalization is slightly higher than for the \cite{bzozo12} parameters (see text) but there is no model parameter adjustment here, at variance with their work. Similar residuals can be obtained for realistic ranges of the three above parameters. Bottom: a small asymmetry is found in all residuals, that we attribute to the fact that the Witte et al longitude is slightly too high compared to the actual flow longitude. A lower longitude should significantly improve the data-model agreement. New studies taking into account the dead time should allow to obtain very accurate flow parameters.}
              \label{figdt}%
    \end{figure*}

For  the study we describe below, we have made an extensive use of the count rates and the model results presented by \cite{bzozo12} in their Fig 18, that is devoted to the year 2009. We have extracted from the Figure the normalized, orbit-integrated data computed in spin angle sectors, as well as the two \textit{old} and \textit{new} model values adapted to the same observing geometry and instrument characteristics. We have concatenated those values to obtain a single series of data for all orbits of interest, and corresponding series of model values. Fig \ref{figdt} shows the concatenated data and the corresponding models, and clearly reveals the trends for the two different models. The \textit{new} model count rates follow the data in a much better way essentially because the rates decrease in the same way as the data as a function of the observer longitude (or orbit number) on both sides of the maximum flux location (the tangential flow location described above). In contrast, the \textit{old} model rates decrease more rapidly than the measurements as the longitude difference increases. The physical reason for a higher or smaller decrease of the fluxes when moving away from the maximum flux location  (valid for measurements made tangentially to the Earth orbit), is the following: the closer the curvature of the observer's orbit (here the Earth orbit) and the curvature of the average He trajectory, the slower the decrease of the entrance atom flux, because the observer motion is better accompanying the helium atom motion (see Fig \ref{figcurve} top). Because most He trajectories in this sidewind location (around longitudes on the order of 130$^{\circ}$) are less bent than the Earth orbit, a slower decrease happens for a lower speed of the atoms at infinity, since their trajectories are more strongly bent at 1 AU. Looking at Fig \ref{figdt} it is clear that this is the dominant effect that favors the low speed found from the IBEX-Lo data (and subsequently the higher longitude, as they are very strongly linked) and disfavors the \textit{old} values. However, this implies that this preferential choice of the low speed is entirely based on count rate variations around the maximum. Any departure between the actual and the hypothesized response to the entrance neutral atom flux may influence the adjustment to the data. It is thus important to consider potential unaccounted for biases in the sensitivity of the count rate to the atom fluxes, since they may influence the derivation of the longitude-speed pair. In what follows we consider the influence of non linear response due to a suppression of events for high counting rates. This effect is very likely to be present and has been mentioned and partly studied by \cite{mobius12}. Our goal is solely to investigate whether the flow parameters found by Witte et al may be compatible with the data when accounting for non-linearity. A full adjustment of parameters is far beyond the scope of this study. 

The existence of high rate suppression of events is documented by \cite{mobius12}. Both the time to record an event by the detector, and the time taken by the Central Electronic Unit (CEU) to analyze the event, construct histograms and communicate with the spacecraft result in a loss of true events. Let N$_{V}$  and N$_{TM}$ be the number of events in a second of counting time of true events, and events transmitted through telemetry respectively. The number of events N$_{TM}$ sent to telemetry is lower than N$_{V}$. This instrumental effect, which results in a non linearity of response, is thoroughly discussed in \cite{mobius12} Appendix A and B.  It can be described in terms of a dead time $\tau_{D}$ during which the instrument cannot record a new event after the proper record of one event. It is well known that the value of N$_{V}$ can be retrieved from N$_{TM}$ if $\tau_{D}$ is known:
\begin{equation}
N_{V}=N_{TM}/(1-N_{TM}*\tau_{D})
\end{equation}

It is clear from the formula that the larger N$_{TM}$, the larger the loss of true events, measured by the factor N$_{V}$/N$_{TM}$. The events are triggered either by the detection of one He atom, or by some residual electrons forming a background with a true rate R$_{BGV}$ \citep{mobius12}. Therefore, N$_{V}$ = N$_{HeV}$ +R$_{BGV}$ and similarly N$_{TM}$ = N$_{Hec}$ +R$_{BGc}$ are the sum of the two types of events, with the index He of Helium atoms events, BG for electrons, V for \textit{true} and c for \textit{counted}. Therefore, equation (1) may be transformed (to let appear explicitly the electron background) into: 
\begin{equation}
N_{HeV}+R_{BGV}=(N_{Hec}+R_{BGc})/(1-(N_{Hec}+R_{BGc})*\tau_{D})
\end{equation}
which may be decomposed in two formulae:
\begin{equation}
N_{HeV}=N_{Hec}/(1-(N_{Hec}+R_{BGc})*\tau_{D})
\end{equation}
that is identical to formula (B1) of \cite{mobius12} and
\begin{equation}
R_{BGV}=R_{BGc}/(1-(N_{Hec}+R_{BGc})*\tau_{D})
\end{equation}
Both the He counts and the background counts are reduced by the same ratio by the dead time effect. It must be kept in mind that, even if R$_{BGV}$ is constant, the \textit{observed} background $R_{BGc}$  is not constant. The following numerical values are taken from \cite{mobius12}. On average, IBEX-Lo counts R$_{BGc}$=22 electrons s$^{-1}$. At its peak, the observed Helium count rate is reaching  $N_{Hec}\simeq$ 25.3 counts s$^{-1}$. \cite{mobius12} indicate in their Appendix B that a dead time of up to 5 ms was derived from their analysis, but that this estimate is not very accurate, owing to the various assumptions made for its derivation. Introducing these values in the above equations, and noting that $N_{HeV}/N_{Hec}=R_{BGV}/R_{BGc}$  (both He atoms and electrons are reduced in the same proportion by the dead time effect), we find $N_{HeV}\simeq$1.31$N_{Hec}$. Interestingly this ratio is on the order of the ratio between the \textit{old} and \textit{new} model values for the second and third orbits before and after the maximum count orbit, i.e. on the order of what differentiates the two models.

\cite{mobius12} have discussed this point, in the context of determining the temperature of Helium by analyzing the variation of the He flow as a function of the spin phase (roll angle about the spin axis) when IBEX was near the maximum of Helium detection. The authors explain in their Appendix B that if $\tau_{D}$ is not negligible, the relation between the observed and \textit{true} rates is a function of the spin angle, and the observed angular distributions appear wider than the original distributions. If uncorrected, this would translate into derivation of an apparently higher temperature. Following these words of caution, \cite{mobius12} consider both the absence and presence of a high rate suppression of events (or dead-time effect) in their determination of the flow temperature. Fig \ref{figcurve} shows one of their synthetic analytical results, namely the temperature-longitude relationships derived from the Earth longitude for maximum flux, and both the maximum flux spin phase and flux peak angular width. It is interesting to see from the Figure how the \textit{old} longitude value of $\simeq$ 75.4 $^{\circ}$ compares with the \textit{old} temperature value of $\simeq$ 6,300K and the \textit{old} speed 26.2 km.s$^{-1}$. Their compatibility requires a dead-time slightly above the value of 5ms, i.e a value they consider as realistic. On the other hand, the \textit{new} longitude value of $\simeq$ 79.2 $^{\circ}$ is compatible with a temperature of $\simeq$6,000K and the \textit{new} speed on the order of 23 km.s$^{-1}$ in the absence of dead-time effect.

In  their extensive comparison of IBEX-Lo data with sophisticated models of interstellar Helium flow at 1 AU, \cite{bzozo12} do not mention any dead time correction, suggesting they have not taken into account its potential effect discussed above in their numerical modeling of the first two years of data. This is in agreement with the compatibility between the longitude, temperature and speed they have derived, that correspond to case of null dead-time, according to the Fig. 8 from \cite{mobius12}. In view of these considerations, we have studied the modifications of the data-model agreement in the frame of dead-time existence, using the data and model values quoted by  \cite{bzozo12}. Our goal is to investigate whether dead time effects may reconcile the  flux decrease predicted by the \textit{old} model with the data when moving away from the maximum flux location. The first argument is that neglecting these effects is reducing the amplitude of the variations, as just explained. The second one is based on the background removal. If, as we assume, \cite{bzozo12} did not take the dead time into account, this implies they have subtracted from the observed count rates a constant background rate of electrons, which value may be derived from the observed data when the Helium flow is expected to be zero (i.e., when the spin angle is away from the helium flow). As we have shown above, in the presence of dead time effects, removing a constant background is not applicable since the counted background RBGc is not constant and depends on the Helium flow count rate, and since both effects conspire to decrease the contrast between high and low He rates when going from true values to observed values.%

We have used for our study a grid of three assumed parameters, i.e. the dead time $\tau_{D}$, the maximum measured value of the Helium count rate N$_{Hec}^{max}$, the background value R$_{BGc}$ removed by \cite{bzozo12}, a value chosen to ensure a decrease to zero of the observed rate far from the helium flow detection. We have varied those parameters within ranges that include the realistic values quoted by \cite{mobius12}. The choice of $\tau_{D}$ and R$_{BGc}$ is imposing the \textit{true} value of the background R$_{BGV}$ based on the formula (4) and N$_{Hec}$=0. For each measurement in Fig \ref{figdt}, that has been normalized at its maximum value over all orbits by \cite{bzozo12}, we multiply this normalized value of the observed count rate by N$_{Hec}^{max}$ to represent the measured rate in counts/s, then add the measured background R$_{BGc}$. This gives the measured rate before background removal. We then transform this rate into a \textit{true} total rate N$_{HeV}$+R$_{BGV}$ (that would have been observed without dead time effect) by using the formula (1). After subtraction of  R$_{BGV}$ we are left with the \textit{true} He rate. We then re-normalize this \textit{true} rate and compute the quadratic sum of differences between this rate and the Witte et al normalized model.  We finally extract from the grid of results those displaying a good agreement between data and model, and for the most realistic parameters, i.e. dead time values on the order of 5 ms, maximum observed flux rates on the order of 25 counts s$^{-1}$ and background on the order of 22 counts s$^{-1}$. Since the methods used by \cite{mobius12} and \cite{bzozo12} differ and we do not know the exact values of the removed background and maximum sector-average flux used in the latter work, we allow for a very small departure from these two parameters.

We show in Fig \ref{figdt} one example of our computations, chosen for its background and maximum count rate values that are very close to the values quoted by \cite{mobius12}, and for a dead time is 7ms, a value that is acceptable according to Fig \ref{figcurve} bottom. We have superimposed the true count rate and the scaled Witte et al model values. We can compare the quality of this solution by comparing the data-model residuals with those derived from the \textit{new} model values taken from \cite{bzozo12}. The sum of squared residuals for this example is 0.0034 to be compared with 0.0031 (using the same normalization), demonstrating that such solutions should definitely be considered, especially considering that we did not perform here an adjustment of the model parameters  and instead kept the Witte et al model.  Smaller residuals could certainly be found if a real adjustment of the model parameters were done. Indeed, interestingly the residuals show a small, but significant increase with the Earth longitude (Fig \ref{figdt} bottom). We believe that this trend demonstrates the need for a slightly lower longitude than the \textit{canonical} one. We note that a longitude closer to 75$^{\circ}$ than 75.4$^{\circ}$ would be in very good agreement with a $\simeq$7ms dead time according to Fig 8 from \cite{mobius12} (or Fig 1 bottom). On the other hand, even this relative difference of 10\% on the residuals is smaller than the distance from the primary to the secondary and unexplained minimum of the $\chi^{2}$ found by \cite{bzozo12}, if we use as a unit the variance around the primary minimum. The good agreement obtained between data and model, based on the Witte et al parameters is a strong argument in favor of the relevance of those parameters, especially given the two extraordinary coincidences mentioned above, namely that the measurements that do not depend on the proportionality of the count rate to the incoming atom flux provide parameters (the latitude and the longitude-speed pair) in very good agreement with those \textit{canonical} values.  Again, a full adjustment of model parameters is beyond the scope of this study, our focus being the temporal variation of the longitude. In this respect, we argue that it is premature to use the \textit{new} longitude and velocity before the question of the dead time correction is fully settled. On the other hand, this first order study shows that future additional IBEX-Lo data analysis should provide extremely precise measurements of the flow parameters.

\section{Other data}
\subsection{Pickup-ions}

\cite{drews12} have analyzed He$^{+}$, O$^{+}$ and Ne$^{+}$  PUI flux measurements made by the PLAsma and SupraThermal Ion Composition instrument (PLASTIC) instrument on the Solar TErrestrial RElations Observatory mission (STEREO)  between 2007 and 2010 \citep{galvin08,drews10}. The four orbits allow the He$^{+}$ focusing cone to be detected four times and the Ne$^{+}$
cone twice. There are also very broad and shallow flux enhancements (compared to the sidewind minima) approximately centered around the upwind direction called \textit{crescent} maxima. The authors first show Gaussian fits to the average cones of He$^{+}$ and Ne$^{+}$ and the O$^{+}$ crescent, superimposed on the measurements. Fluxes are averaged over the four orbits. They use a statistical method to estimate the error on the centers of those three features, instead of errors generated during the fitting procedure. The best-fit flow longitude values they  quote are 75.0$\pm0.3^{\circ}$, 75.0$\pm2.9^{\circ}$ and 79.2$\pm1.4^{\circ}$ for He$^{+}$ and Ne$^{+}$ and O$^{+}$  respectively.  The first two values are in good agreement with the \textit{consensus} longitude value. The third is out of the allowed range. 
In a second step of the analysis they show a sophisticated, extensive statistical study of the whole dataset based on the assumption that the flow direction has not changed over the four years of data, and that the ionization of the neutral atoms responsible for the PUIs has varied very smoothly. As a consequence, the very strong fluctuations of the observed PUI fluxes can be decoupled from their creation and result only from the solar wind variability. The output of this study is a series of eleven values for the flow longitude derived from the eleven possible combinations of orbits. Based on this statistical study they derive weighted mean values of the  flow longitude of 77.4$\pm1.9^{\circ}$ and 80.4$\pm5.4^{\circ}$ from the He cone and \textit{crescent}, 77.4$\pm5.0^{\circ}$ and 79.7$\pm2.6^{\circ}$ from the Ne cone and \textit{crescent}, and finally 78.9$\pm3.1^{\circ}$ from the O$^{+}$ \textit{crescent}. 

The question we are addressing in this study is the existence of a temporal variability of the interstellar flow outside the heliosphere, and which measurements can be used for this purpose. This includes: -(i) which type of method is likely to provide accurate values of the flow within the heliosphere?, -(ii) which species are appropriate for the derivation of the flow outside the heliosphere?, -(iii) are PUI fluxes maxima appropriate in general for the direction of the flow? Hereafter we consider these three points. 

(i) Although it is ingenious and able to provide estimates of errors, we have concerns about the method based on orbit combinations and the separation of temporal and spatial effects. The main concern is the assumption about the \textit{smoothly} varying ionization. \cite{drews12}  consider only the photoionization, and omit the second source of ionization, electron impact. Theoretical calculations and observational studies of the neutral helium cone with UVCS on SOHO have demonstrated that electron impact can be responsible for a very large fraction of PUIs, and, important here, especially in response to dense solar wind streams that possess large supra-thermal electron tails \citep{ruci89,lalluvcs}. According to this study, in stationary conditions and at 1 AU the fraction of PUIS born after electron impact may reach 30\% of their total flux, both on the upwind and downwind sides, and this effect may explain temporarily observed correlations between H and He PUIs fluxes, as well as the anti-correlation between the wind speed and the He$^{+}$ fluxes. This means that in non-stationary conditions there may be a very strong variability from one region to the other in response to solar streams as well as local ratios between the two kinds of PUIs significantly above this value. This also implies that the assumed absence of coupling between the solar wind induced variability and their creation by ionization of neutrals is not a fulfilled condition. We have also other concerns about the assumption of eleven fully independent determinations, since the combinations or orbits are based on four orbits only, and thus the eleven realisations are not fully independent. As a conclusion , we believe that the direct Gaussian fits to actual data mentioned above provide more reliable values of the PUI flux maxima longitudes.

(ii) The use of oxygen ions for the purpose of deriving the direction of the oxygen flow that enters the inner heliosphere is appropriate, but not for the derivation of the flow outside the heliosphere. As  a matter of fact, as quoted by \cite{drews12}, a deflection of the flow due to secondaries created by charge-exchange is a likely as for hydrogen. Subsequently, the O$^{+}$ crescent should not be used for the interstellar flow longitude study and should not be included in the \cite{crim13} study. 

(iii) As mentioned by \cite{drews12}, transport effects may be significant for PUIs. \cite{chalov06} have computed such effects and find potential deviations on the order of 5$^{\circ}$ on the He$^{+}$ downwind cone at 1 AU. Interestingly, due to the Parker spiral orientation, the PUIs maxima are always displaced towards {\bf higher} longitudes. The amplitude of the displacement depends on the level of turbulence, the SW characteristics and the distance to the Sun. As a consequence the average longitudes of the PUIs enhancements should be considered as {\bf upper limits} of the interstellar neutral flow longitude. Indeed, one very interesting result of \cite{drews12} is the systematic and strong discrepancy they find between the longitudes derived from the upwind and downwind data, that in principle should be identical. The authors mention the transport effects as the likely source of this discrepancy, reinforcing our argument that is not possible to use the PUI-derived longitudes in another manner than as upper limits. We suggest that the upwind deviations are likely to be stronger than for the focusing cones, since PUIs are created closer to the Sun on the upwind side (see e.g. Fig 11 of \cite{lalluvcs}, thus upwind PUIs collected at 1 A.U. may have experienced stronger turbulence and more transport effects compared to the downwind PUIs. The \cite{drews12} \textit{crescent} and cone discrepancies certainly deserve further study, with the transport effects taken into consideration. Importantly, diffusion along the field lines may be much stronger in 2007-2010 at a time of very weak activity, compared to the conditions prevailing in 2000 and experienced by the PUIs recorded by ACE \citep{gloeckler04}.

We conclude from (i) to (iii) that only the \cite{drews12} directly fitted average He$^{+}$ and Ne$^{+}$ cones can be used, and only as upper limits on the flow longitude.

\subsection{Solar backscatter radiation}

\cite{crim13} include in their analysis a flow direction corresponding to the Mariner 10 analysis \citep{ajello78, ajello79}. We note however that the direction quoted in this work has been derived from H Ly$\alpha$ observations, namely from the downwind Ly$\alpha$ minimum, i.e. it applies to the neutral hydrogen and not to helium. In the early years of UV glow measurements it had not been considered that the helium and hydrogen flows could enter the solar system from different directions, and the authors had adjusted models to the Mariner 10 He and H data for the same directional parameters. Since then measurements are much more  precise and a  difference could be  established \citep{lall05}, and indeed the longitude of the hydrogen flow has been found smaller by about $\simeq2-3^{\circ}$ than the helium longitude. Consequently it does not seem appropriate to use this direction in the context of the helium direction.

\cite{crim13} also include in their analysis two distinct results based on the Prognoz satellite measurements of 58.4 nm backscattered radiation, a first longitude determination of 72.2$^{\circ}$ and a second one of 74.5$^{\circ}$. The first lower value has some influence on their inferred positive slope for the evolution of the flow longitude. However, a careful look at the \cite{dalo84} article  shows clearly that the first value should definitely {\bf not} be included. \cite{dalo84} present this first determination as a preliminary check (based on angles roughly determined from a fraction of the scans) that their dataset well corresponds to the helium cone (this was the first analysis of the Prognoz helium data) and do not consider further this result, which is not even mentioned in the abstract nor in the conclusion. More importantly, the authors quote the value of 72.2 as an {\bf eye-fit} preliminary determination based on the fraction of the data and it is clear from the whole article that it can by far not compete with the second result, based on the whole dataset and a careful data-model adjustment. We conclude that the use of this first value is not justified.

\cite{crim13} also include the results of \cite{hiromu08} obtained on board Nozomi. They use for the longitude an uncertainty of 3.4$^{\circ}$. A careful look at the article reveals that: (i) there is no justification anywhere of such an uncertainty, e.g. there is no mention nor presentation of a data-model adjustment, (ii) there is not a single figure showing any comparison of the helium data (or even a fraction of the data) with any model allowing this estimate to be convincing, and (iii) the article is entirely devoted to H-Lyman alpha while the helium analysis is presented in a three line paragraph! From looking at the Figures 6 and 7 showing the He measurements, that are done with a pixel angular resolution of $\simeq6^{\circ}$, it is clear that the error bar of 3.4$^{\circ}$ on this longitude estimate is underestimated. From the data shown in Figure 6 and 7 we estimate that the actual error is at least on the order of $\simeq6^{\circ}$, the size of the angular pixel. In what follows we have used this value for the error.

   \begin{figure*}
   \centering
\includegraphics[width=\linewidth]{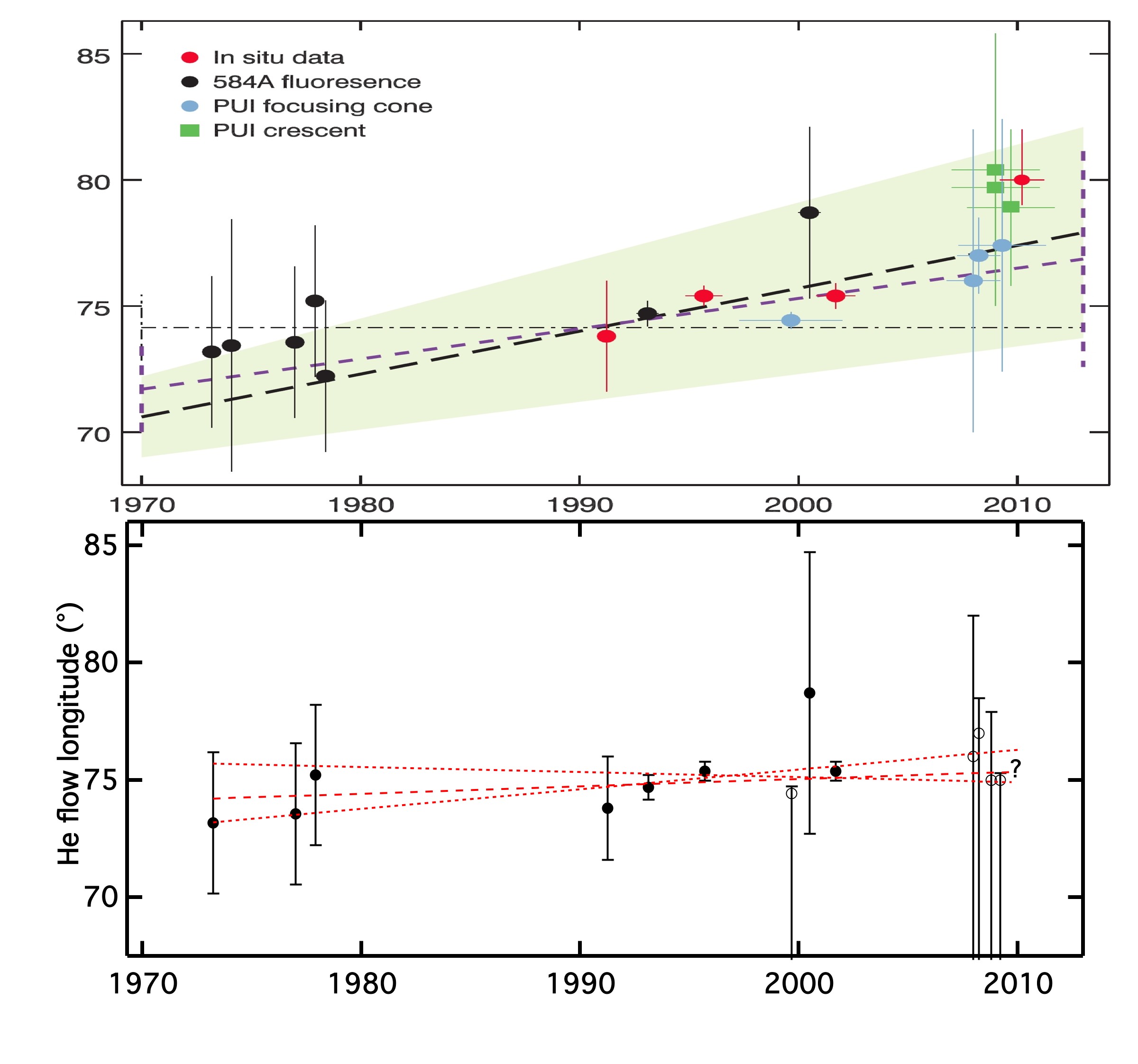}
   \caption{Top: the original figure from \cite{crim13} showing the inferred temporal increase of the helium wind longitude. Bottom: our revisited set of appropriate measurements.  Filled circles are the data points we consider for linear fits. Longitudes derived from PUI data (augmented by 2 $\sigma$) serve as upper limits only. Shown are the central, upper and lower trends allowed by the fits. Stationarity is contained within  the solutions. The question mark corresponds to the IBEX period and the Witte et al longitude, not included in the fit (the IBEX-Lo result is also not included, according to Section 2). PUI data points are slightly displaced on the Figure to avoid overlap.}
              \label{figtemporal}%
    \end{figure*}

\section{Synthesis and conclusion}

We have revisited the series of measurements used by \cite{crim13} to infer a temporal increase of the helium flow longitude over the last decades. 
Motivated by the fact that both the flow latitude and the combination of speed and longitude deduced from IBEX-Lo count rate maxima are very surprisingly fully compatible with the canonical set of parameters derived in 2004, and that the preferential choice of a lower speed and higher longitude depends essentially on the count rate response to the atom flux, we have examined the potential consequences of high rate suppression of events (dead time effects) likely present in the IBEX data \citep{mobius12}. The study shows that the \textit{old} set of parameters, including a flow longitude of 75$^{\circ}$, is as likely as the \textit{new} one, provided a dead time of the order of 7ms is included, i.e. at a level estimated to be realistic by these authors. We conclude that the IBEX-Lo parameters can not be used for a temporal study until such effects are further examined. Based on the two mentioned coincidences it is very likely that the \textit{old} set of parameters (and even probably a slightly lower longitude), is indeed the actual solution. 

Based on published models \citep{chalov06} we argue that PUI measurements can provide only an upper limit on the flow longitude, due to transport effects. At variance with ACE data in 2000 \citep{gloeckler04}, the STEREO data were recorded at a time of particularly low activity, which favors the diffusion along the magnetic field lines, and potentially produces displacements of the PUI average trajectories by a up to 5$^{\circ}$, according to \cite{chalov06}. In this respect it is particularly interesting that the upwind maxima are systematically shifted to higher longitudes compared to the downwind focusing cones, as noted by \cite{drews12}.  We also argue that the statistical study of the STEREO PLASTIC data is not adequate, being based on the assumption of dominant and smoothly varying ionization by UV photons. This assumption is highly questionable, since a large fraction of helium ions originates from electron impact ionization, especially within the Earth orbit, and this electron impact ionization is  a highly variable effect linked to dense solar wind streams and their suprathermal electron  tails \citep{lalluvcs}. 

We have also argued that the use of a fraction of the backscatter measurements and errors is inappropriate, in particular the flow direction from \citep{ajello78,ajello79} that is derived from H Ly$\alpha$ data and the \textit{eye-fit} preliminary direction of \cite{dalo84}. We also revisited the error bar for Nozomi Helium data \citep{hiromu08}. 

Fig \ref{figtemporal} displays the various longitude measurements and corresponding errors we believe can be used at present, according to the above remarks. Longitudes derived from PUI data (augmented by 2 $\sigma$) serve as upper limits only. Actually one PUI measurement, the 2000 ACE data \citep{gloeckler04} is adding a constraint. A weighted linear fits to those data shows no signs of temporal increase (including or not the PUIs upper limits). This stability is in agreement with the absence of variation between the early value of the H flow direction (e.g. $\lambda$=72$^{\circ}$ from \cite{ajello78}) and the most recent one with SOHO/SWAN \citep{lall10}. Further analyses of the whole set of IBEX-Lo data at high temporal resolution, and taking into account as carefully as possible dead-time correction and electron noise should provide the most accurate parameters on the He flow among all experiments, allowing better inter-comparisons and determinations of temporal trends, if they do exist. 

\begin{acknowledgements}
We deeply thank our referee for his careful reading and numerous and useful suggestions.
This work has been funded by CNRS and CNES grants.
\end{acknowledgements}


\end{document}